\documentclass[a4paper,twoside]{article}
\usepackage{verbatim}
\usepackage{algorithmicx}
\usepackage{algorithm}
\usepackage{algpseudocode}
\usepackage{changebar}

\usepackage{amsthm}
\usepackage{amssymb}
\usepackage{amsmath}
\usepackage{amsfonts}
\usepackage{rotating}
\usepackage{stfloats} 
\usepackage{subfig}
\usepackage{eqparbox}
\usepackage{float}
\usepackage{graphicx,url}
\usepackage{todonotes}
\usepackage{version}
\usepackage{setspace}	
\usepackage{array}
\usepackage{pslatex}
\usepackage[compact]{titlesec}

\usepackage{cite}
\usepackage{apalike}

\usepackage{SCITEPRESS}     

\newtheorem*{theorem*}{Theorem}
\newtheorem{constraint}{Constraint}
\newtheorem{definition}{Definition}
\newtheorem{lemma}{Lemma}

\makeatletter

\renewcommand\section{\@startsection {section}{1}{\z@}%
                                   {-2ex \@plus -1ex \@minus -.2ex}%
                                   {+0.5em}%
                                   {\normalfont\large\bfseries}}
\renewcommand\subsection{\@startsection {subsection}{1}{\z@}%
                                   {-2ex \@plus -1ex \@minus -.2ex}%
                                   {+0.5em}%
                                   {\normalfont\normalsize\bfseries}}
\makeatother

\begin{document}

\title{On the Evaluation of the Privacy Breach in Disassociated Set-Valued Datasets}

\author{\authorname{Sara Barakat\sup{1}, Bechara Al Bouna\sup{1}, Mohamed Nassar\sup{2}, and Christophe Guyeux \sup{3}}
\affiliation{\sup{1}TICKET Lab., Antonine University, Lebanon}
\affiliation{\sup{2}Department of Computer Science, MUBS, Lebanon}
\affiliation{\sup{3}University of Bourgogne Franche-Comt\'e, France}
\email{\{sara.barakat, bechara.albouna\}@ua.edu.lb, meb.nassar@gmail.com, christophe.guyeux@univ-fcomte.fr\vspace*{-30pt}}
}

\keywords{disassociation, privacy breach, data privacy, set-valued, privacy preserving}

\abstract{
Data anonymization is gaining much attention these days as it provides the fundamental requirements to safely outsource datasets containing identifying information.  While some techniques add noise to protect privacy others use generalization to hide the link between sensitive and non-sensitive information or separate the dataset into clusters to gain more utility. In the latter, often referred to as bucketization, data values are kept intact, only the link is hidden to maximize the utility. 
In this paper, we showcase the limits of disassociation, a bucketization technique that divides a set-valued dataset into $k^m$-anonymous clusters. We demonstrate that a privacy breach might occur if the disassociated dataset is subject to a cover problem. 
We finally evaluate the privacy breach using the quantitative privacy breach detection algorithm on real disassociated datasets. 
\vspace*{-16pt} 
}
\onecolumn \maketitle \normalsize \vfill
\section{Introduction} \label{sec:introduction}

Privacy preservation is an important requirement that must be considered carefully before the release of datasets containing valuable information. Anonymizing a dataset by simply removing the personally identifying information (PII) is insufficient to prevent a privacy breach \cite{kanon_defn,kanon_sweeney}. An attacker\footnote{a person who is intentionally trying to link individuals to their sensitive information} may still succeed in associating his/her background knowledge with one or multiple records via non-sensitive information in the dataset to eventually reveal individuals' sensitive information. The same problem arises when releasing set-valued data (e.g., shopping and search logs) consisting of a set of records where each record links an individual to his/her set of distinct items. 
The AOL search data leak in 2006 \cite{aol2006} is an explicit example that shows the implications of inappropriate anonymization on data privacy. The query logs of 650k individuals were released after omitting all explicit identifiers. They were later withdrawn due to multiple reportings of privacy breaches. 

\begin{figure}[htp]
	\centering
\subfloat[Original set-valued dataset]{
		\includegraphics[width=0.47\textwidth]{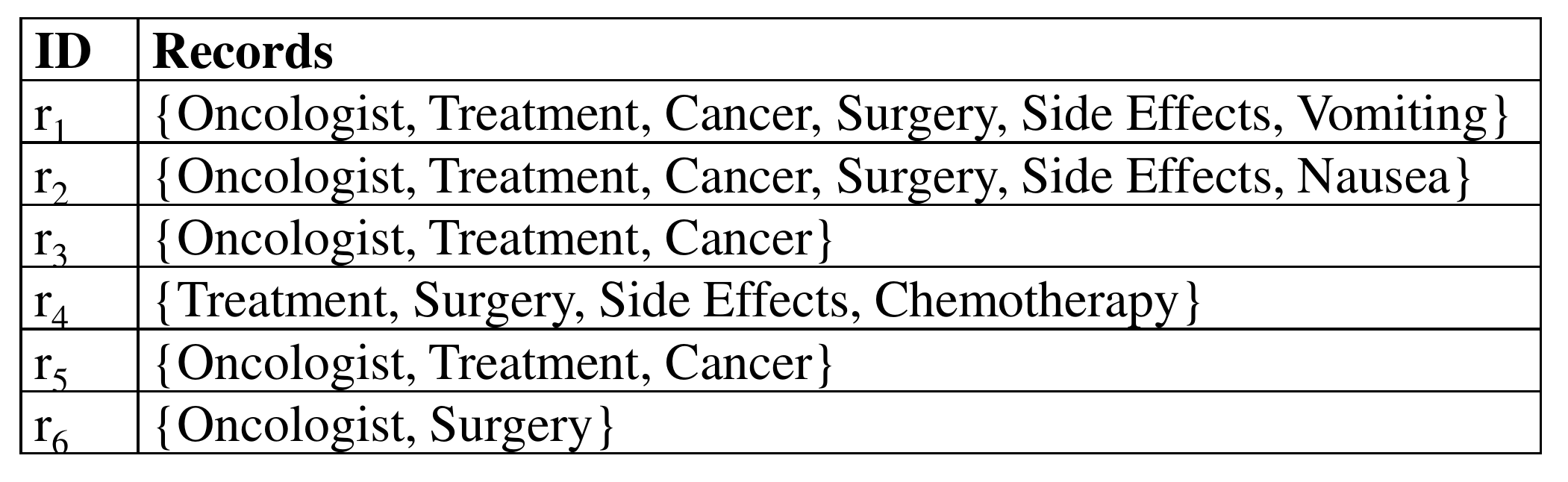}
	\label{fig:original-a}
}\\  
\subfloat[Disassociated dataset with record chunks $C_1$ and $C_2$]{
		\includegraphics[width=0.47\textwidth]{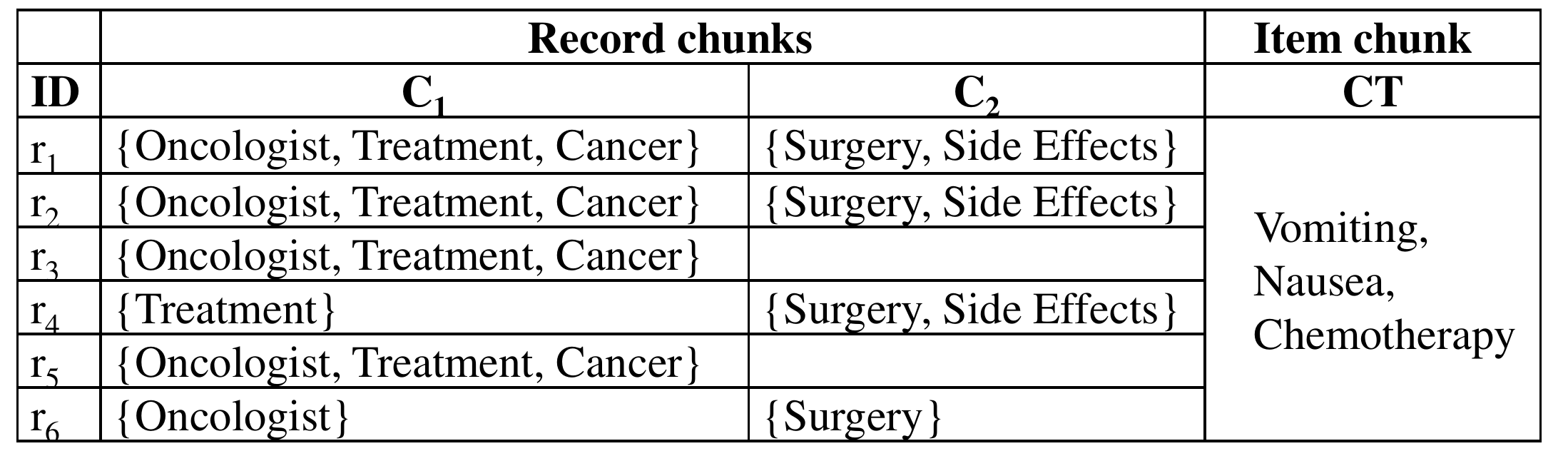}
	\label{fig:disasso-b}
}
\caption{Example scenario }
\label{fig:setvalued}
\end{figure}

To better illustrate the problem, let us consider a dataset of individuals and their corresponding sets of searched data items. Let us assume that an attacker knows that Alice, an individual whose items figure in publicly available dataset, has used two search items: \{$Side\, Effects$ and $Vomiting$\}. The attacker's background knowledge consists of the set of these two items. If it happens that one and only one record exists in the released dataset where both $Vomiting$ and $Side\, Effects$ belong to the same itemset, such as the case of record $r_{1}$ in the example shown in Figure \ref{fig:setvalued}, the attacker can link the individual Alice to $r_{1}$. Here, the attacker, who has partial knowledge of some of the data items searched by the individual, is able to determine the complete itemset and link it to the corresponding individual.

There has been significant work in anonymization \cite{kanon_defn,kanon_sweeney,ldiversity,anatomy,Dwork2006,terrovitis2012} to cope with this problem. Categorization, on the one hand, separates the data into sensitive and non-sensitive categories of values, which unfortunately can be hard to adopt in real world applications due to the elastic meaning of the sensitivity \cite{terrovitis2012}. Generalization, on the other hand, transforms values into a general form and creates groups to eliminate possible linking attacks. Despite its efficiency in privacy preservation, generalization has a major drawback related to the loss in data utility \cite{anatomy}, making it less efficient for aggregate analysis. Anonymization by disassociation \cite{terrovitis08,loukidesLGT14,Loukides2015} is one recent method that preserves the original items keeping them intact, without applying generalization or substitution. Instead, it uses some form of bucketization and separates the records into clusters and chunks to hide their associations. More specifically, disassociation transforms the original data into $k^m$-anonymous \cite{terrovitis08} records by dividing the dataset 1) horizontally, creating clusters of similar records and 2) vertically, creating $k^m$-anonymous chunks of similar sub-records (e.g., record chunks $C_1$ and $C_2$ in Figure \ref{fig:disasso-b}) and an item chunk containing items that appear less than $k$ times.


Disassociation lies under a general category of techniques  \cite{slicing,anatomy,fragmentation} that preserve privacy by dividing the dataset to maximize the utility. It provides strong theoretical guarantees but unfortunately remains vulnerable when the dataset is subject to what we call a \textit{cover problem}, which is defined in Section~\ref{sec:deanon}.


In this article, we evaluate the privacy provided by the disassociation of a set-valued dataset and demonstrate that it might be breached whenever this dataset is subject to a \textit{cover problem}. We show that attackers with different levels of partial background knowledge weak, moderate, or strong are able to compromise the privacy of a disassociated dataset. We also show that, as many other bucketization techniques \cite{anatomy,slicing} in their attempt to protect against attribute disclosure, preserving privacy tends to be overly optimistic \cite{correlation,definetti,bechara14,bechara15}. 


The contributions of this research work can be summarized as follows:

\begin{itemize}
 \item We present a formal model of the set-valued data and the disassociation technique.
\item We define the cover problem and investigate its repercussion on the disassociation of the dataset.
 \item We propose the so-called quantitative privacy breach detection algorithm, and study its efficiency. The proposed algorithm shows to what extent the privacy of a disassociated dataset can be breached. 
  

\end{itemize}
The remainder of this paper is organized as follows. In the next section, we present prior work in data anonymization and discuss some of the techniques used to breach well known privacy constraints. In Section \ref{sec:dissassociation}, we give some insights on privacy preserving using the disassociation technique. Section \ref{sec:attackers} defines the attacker model including the three types of attackers: strong, moderate, and weak. In Section \ref{sec:deanon}, we present the cover problem and the resulting privacy breach in a disassociated dataset. We finally present a set of experiments to evaluate the de-anonymizing algorithm is presented in Section \ref{sec:experim}. 

\section{Related Work} \label{sec:relwork}

\textbf{Anonymization techniques} have been a topic of intense research in the last decade. 
Several have been proposed as a way to protect privacy by guaranteeing that the probability of linking an individual to a specific record will not be greater than a certain threshold (e.g., $1/k$ or $1/l$). Some rely on generalization \cite{kanon_defn,kanon_sweeney,ldiversity}; replacing some values with more general values based on a given  taxonomy or an interval for numerical values, while others on bucketization \cite{anatomy,fragmentation,fragmentationInference}; separating what is sensitive from non-sensitive. 
Disassociation lies under this category of bucketization techniques, preserves utility by keeping the values intact without any alteration, but it does not divide the attributes into sensitive and non-sensitive. 

 Dwork defined the notion of differential privacy \cite{Dwork2006} to handle private data publishing efficiently. It gained much popularity among computer scientists providing strong assumptions on the way that data should be released. It is based on a strong mathematical foundation and guarantees that an attacker will not be able to learn any information about an individual that he/she already knows if the individual is removed from the dataset. 
 In order to achieve differential privacy, approaches tend to release privatized data by introducing noise to query results. Here, despite the efficiency provided by differential privacy, we opt for disassociation since it publishes truthful datasets. 

Yeye et al. \cite{yeye2009} extended $k$-anonymity
to set-valued context, in which there is no distinction between quasi-identifying, sensitive or non-sensitive
values. In their approach, the authors generate a $k$-anonymous dataset via a top-down partitioning
privacy technique based on local generalization. 
In a similar approach, the authors in \cite{fard2010effective} adopted the $k$-anonymity privacy constraint for a transactional dataset. They proposed a clustering based technique
to group \textit{similar} transactions into clusters, in order to reduce generalizations and suppression thus reducing information loss. A common problem with these methods is that a large part of the initial items are usually missing from the anonymized data due
to generalization or suppression. 
In \cite{Jia2014}, the authors propose a privacy notion $\rho$-uncertainty to ensure
that there are no sensitive association rules that can be inferred with confidence
higher than $\rho$. Truthful association rules however can still be
derived, but this requires, in a similar manner to bucketization techniques, distinction between the sensitive and non-sensitive values.

Preserving privacy by disassociation is a promising technique in which the values are kept intact without any alteration: neither generalization nor suppression. It thus falls under the general category of bucketization techniques but again, without the need to distinguish between sensitive and non-sensitive values.

\textbf{Attacks on anonymization} focus on the identification of sensitive information, information that is meant to be private, by either exploiting defects in the privacy constraints, a bug in the mechanism that implements it or even by assuming knowledge of the underlying anonymization algorithm such as the minimality attack \cite{cormodeMinimality,wong2007}.\\
We discuss here some of the interesting attacks that have been reported in the data anonymization literature.
Homogeneity attack \cite{kanon_defn} is considered successful when there is a lack of diversity among the values of the sensitive attribute. Background knowledge \cite{ldiversity} attack compromises anonymization, when the attacker's background knowledge includes information about the sensitive values of a specific individual or partial knowledge of the distribution of sensitive and non-sensitive values. 
Unsorted matching attack \cite{Sweeney01computationaldisclosure} is related to the order in which the tuples appear in the released table that can leak sensitive information if it is preserved in the anonymized dataset.\\
The attacks mentioned previously present some insights on the privacy problems that might be encountered when anonymizing sensitive information in an outsourcing scenario. To the best of our knowledge, none has explored the privacy breach in a disassociated dataset. In this paper, we highlight and evaluate the ability of an attacker to link his/her partial background knowledge represented by the $m$ items that he/she is allowed to have, according to the privacy constraint $k^m$-anonymity, to less than $k$ records. 
The works described in \cite{correlation,definettitheorem} are good examples of attacks that follow the same convention by identifying flaws in the anonymization techniques allowing attackers to use their background knowledge to breach privacy or extract foreground knowledge. The authors assume that the correlation between the values in an anonymized dataset could be used to breach the privacy of the anonymization techniques that use $l$-diversity in their underlying privacy mechanism. 

\section{Preserving Privacy by Disassociation} \label{sec:dissassociation}

\subsection{Formalization} \label{sec:formalization}

Let $\mathcal{D} = \{x_1,..., x_d\}$ be a set of items (e.g., supermarket products, query logs, or search keywords). Any subset $I \subseteq \mathcal{D}$ is an itemset (e.g., items searched together). Let $\mathcal{T}=(r_1,..., r_n)$ be a dataset of records, where $r_i \subseteq \mathcal{D}$ for $1\leq i \leq n$ is a record in $\mathcal{T}$ and $r_i$ is associated with a specific individual of a population. We use $[r_i]_m$ to denote the set of all $m$-combinations of the record $r_i$. Equivalently, $[r_i]_m$  is the set of $m$ items subsets of $r_i$.
Note that $\mathcal{D}$ is no more than the set of items in $\mathcal{T}$: $\mathcal{D}=\bigcup^{n}_{i=1}r_i$.


We use $R_\mathcal{T}$ to define a subset of records in $\mathcal{T}$. 
 $s(\mathcal{T})$ and $s({R_\mathcal{T}})$  denote the numbers of records in $\mathcal{T}$ and $R_\mathcal{T}$ respectively and thus, $s(I, \mathcal{T})$ denotes the number of records in $\mathcal{T}$ that contain $I$.

Table \ref{tab:notations} defines the basic concepts and notations used in the paper. 
\begin{table}
\tiny
\centering
\caption{Notations}

\begin{tabular}{|l|p{5cm}|}\hline
$\mathcal{T}$ & a table containing individuals related records \\ \hline
$r$ &  a record of $\mathcal{T}$; set of items associated with a specific individual of a population \\ \hline
$[r_i]_m$ & a set of all $m$-combinations of items in the record $r_i$ \\ \hline
$I$ &  an itemset of $\mathcal{D}$ that might or not be grouped together in a record of $\mathcal{T}$\\ \hline
$R_\mathcal{T}$ &  a set of records in $T$ grouped as a cluster\\ \hline
$s(I, \mathcal{T})$ & the number of records in $\mathcal{T}$ that contain $I$ \\ \hline
$\mathcal{T}^*$ & a table anonymized using the disassociation technique\\ \hline 
$C$ &  a chunk in a disassociated dataset; is a set of sub-records of $\mathcal{T}$ in $\mathcal{T}^*$  \\ \hline
$C_T$ &  an item chunk in  a disassociated; dataset a quasi-identifier group \\ \hline
$R_C$ & a record chunk; is a set of sub-records of $R$ in $\mathcal{T}^*$  \\ \hline
$\mathcal{B}$ & the background knowledge of an attacker \\ \hline
\end{tabular}
\label{tab:notations}
\end{table}

\subsection{Disassociation} \label{sec:disass}

Disassociation works under the assumption that the items should neither be altered, suppressed, nor generalized,  but at the same time the dataset should serve the $k^m$-anonymity privacy constraint \cite{terrovitis08}. $k^m$-anonymity guarantees that an attacker who knows up to $m$ items, grouped together in an itemset $I \subseteq \mathcal{D}$, cannot associate them with less than $k$ records from $\mathcal{T}$, where $k\geq 2$ is a user-defined constant. Formally, $k^m$-anonymity is defined as follows:

\begin{definition}[$k^m$-anonymity] Given a dataset of records $\mathcal{T}$ whose items belong to a given set of items $\mathcal{D}$, we say that $\mathcal{T}$ is $k^m$-anonymous if $\forall I \subseteq \mathcal{D}$ and $|I|\leq m$, the number of records that contain $I$ in $\mathcal{T}$ is greater than or equal to $k$, $s(I, \mathcal{T}) \geq k$.

\end{definition}

In a practical scenario, $k^m$-anonymity cannot be achieved without a severe loss in data utility. 
The disassociation technique \cite{terrovitis2012,Loukides2014}, for its part, ensures privacy through bucketization to provide better utility when it comes for frequent itemsets discovery and aggregate analysis. 

Disassociation separates the dataset into clusters of $k^m$-anonymous record chunks and an item chunk. The key idea is to sort records based on the most frequent items and then group them horizontally into smaller disjoint clusters $\{R_1,...,R_n\}$. It partitions, in a next step, the clusters vertically into $k^m$-anonymous record chunks $\{C_1,...,C_n\}$ and an item chunk $C_T$ to hide infrequent combinations. The record chunks are created subsequently as long as there are items that can be grouped together in a way to satisfy the $k^m$-anonymity privacy constraint. The remaining items, the ones that occur less that $k$ times, are moved to the item chunk $C_T$.\footnote{The shared chunk as defined in the original paper \cite{terrovitis2012} is omitted here for simplicity.}

Formally, given a dataset $\mathcal{T}$, applying disassociation on $\mathcal{T}$ will produce an anonymized dataset $\mathcal{T}^*$ composed of $n$ clusters where each is divided into a set of record chunks and an item chunk,
\begin{equation*}
\mathcal{T}^*=(\{R_{1_{C_1}},..., R_{1_{C_t}},R_{1_{C_T}}\} , ..., \{R_{n_{C_1}},..., R_{n_{C_t}},R_{n_{C_T}} \} )
\end{equation*}
such that $\forall R_{i_{C_j}} \in \mathcal{T}^*$, $R_{i_{C_j}}$ is $k^m$-anonymous,
where,
\begin{itemize}
\item $R_{i_{C_j}}$ represents the itemsets of the $i^{th}$ cluster that are contained in the record chunk $C_j$. 
\item $R_{i_{C_T}}$ is the item chunk of the $i^{th}$ cluster containing items that occur less than $k$ times.
\end{itemize}

The example in Figure \ref{fig:disasso-b} shows that the disassociated dataset contains only one cluster with two $k^m$-anonymous record chunks ($k$=3 and $m$=2) and an item chunk, thus, $\mathcal{T^{*}}$=(\{$R_{1_{C_{1}}}, R_{1_{C_{2}}}, R_{1_{C_{T}}}$\}).

In a disassociated dataset,  privacy is preserved by assuming that the record chunks are $k^m$-anonymous and that the records that are not $k^m$-anonymous in the original dataset, are $k^m$-anonymous in \textit{at least} one of the datasets resulting from the inverse transformation of the disassociated dataset. This privacy guarantee is formally expressed as follows:
\begin{definition}[Disassociation Guarantee]
Let $\mathcal{G}$ be the inverse transformation of $\mathcal{T}^*$. $\mathcal{G}$ takes an anonymized dataset $\mathcal{G}(\mathcal{T}^*)$ and outputs the set of all possible datasets $\{\mathcal{T}'\}$. $\forall I \subseteq \mathcal{D}$, $|I| \leq m$, $\exists \mathcal{T}' \in \mathcal{G}(\mathcal{T}^*) $ where $s(I, \mathcal{T}') \geq k$.  
\end{definition} 

The Disassociation Guarantee ensures that for any individual with a complete record $r$, and for an attacker who knows up to $m$ items of $r$, at least one of the datasets reconstructed by the inverse transformation has $r$ hidden by repeating its existence $k$ times. 

 That is, given $m$ items of $r$, if $r$ exists less than $k$ times in the original dataset, the attacker will not be able to link back the record to the individual with a probability greater than $1/k$ in the disassociated dataset. This record $r$, as all other records, exists in at least one of the inverse transformations $k$ times.   

\section{Attacker Model}\label{sec:attackers}

In accordance with the work in \cite{terrovitis2012,Loukides2014,terrovitis08}, we consider that the attacker intends to link individuals to their anonymized records (e.g., complete set of search logs or purchased supermarket products). This can be done, in a trivial anonymization, by tracing unique combinations of items. For example, an attacker who knows that Alice has searched for side effects and vomiting is represented by a background knowledge set of items \{\textit{Side Effects, Vomiting}\}. Since one and only one such record exists in the dataset of Figure \ref{fig:original-a}, the attacker can easily link this record back to Alice. In addition, we assume that the attacker 
has knowledge of the disassociation technique. 

We also assume that the attacker may have background knowledge consisting of at most $m$ items; knowing these $m$ items, he/she will not be able to link them to less than $k$ records in the dataset (i.e., adopting $k^m$-anonymity, the same privacy guarantee as in \cite{terrovitis2012}).
Some models add to it negative background knowledge (e.g., the attacker knows that an individual has not posed a specific query, that an item is not purchased by a given individual, etc.). Again, as in \cite{terrovitis2012} and many other bucketization techniques \cite{anatomy} \cite{slicing}, we do not assume this kind of negation knowledge. Here, we classify attackers into three categories: strong, moderate, and weak, depending on their level of background knowledge and consequently their ability to associate individuals with their records in $\mathcal{T}$.

\begin{figure*}[ht]
	\centering
	\subfloat[Strong attackers]{
		\includegraphics[width=0.24\textwidth]{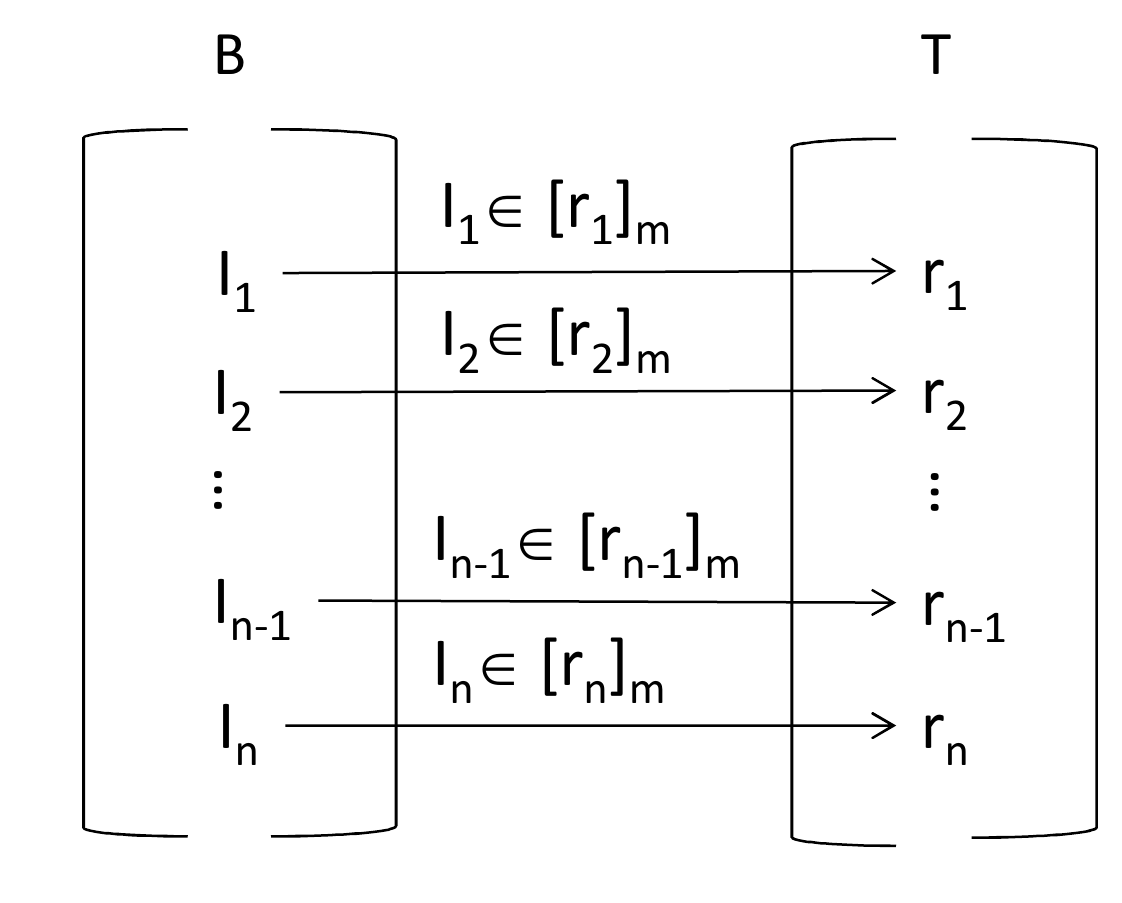}
\label{fig:strong}
	}
\subfloat[Moderate attackers]{
		\includegraphics[width=0.26\textwidth]{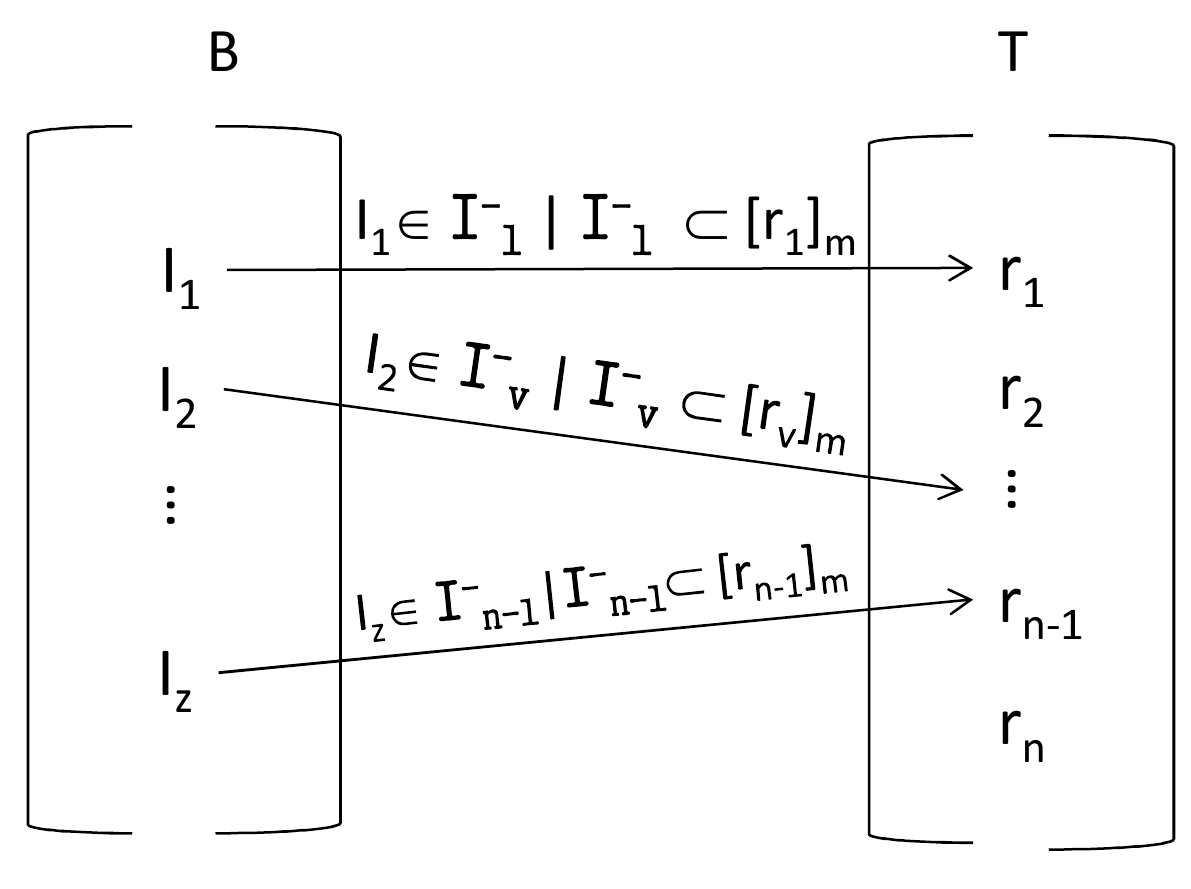}
	\label{fig:moderate}
}
\subfloat[Weak attackers]{
		\includegraphics[width=0.26\textwidth]{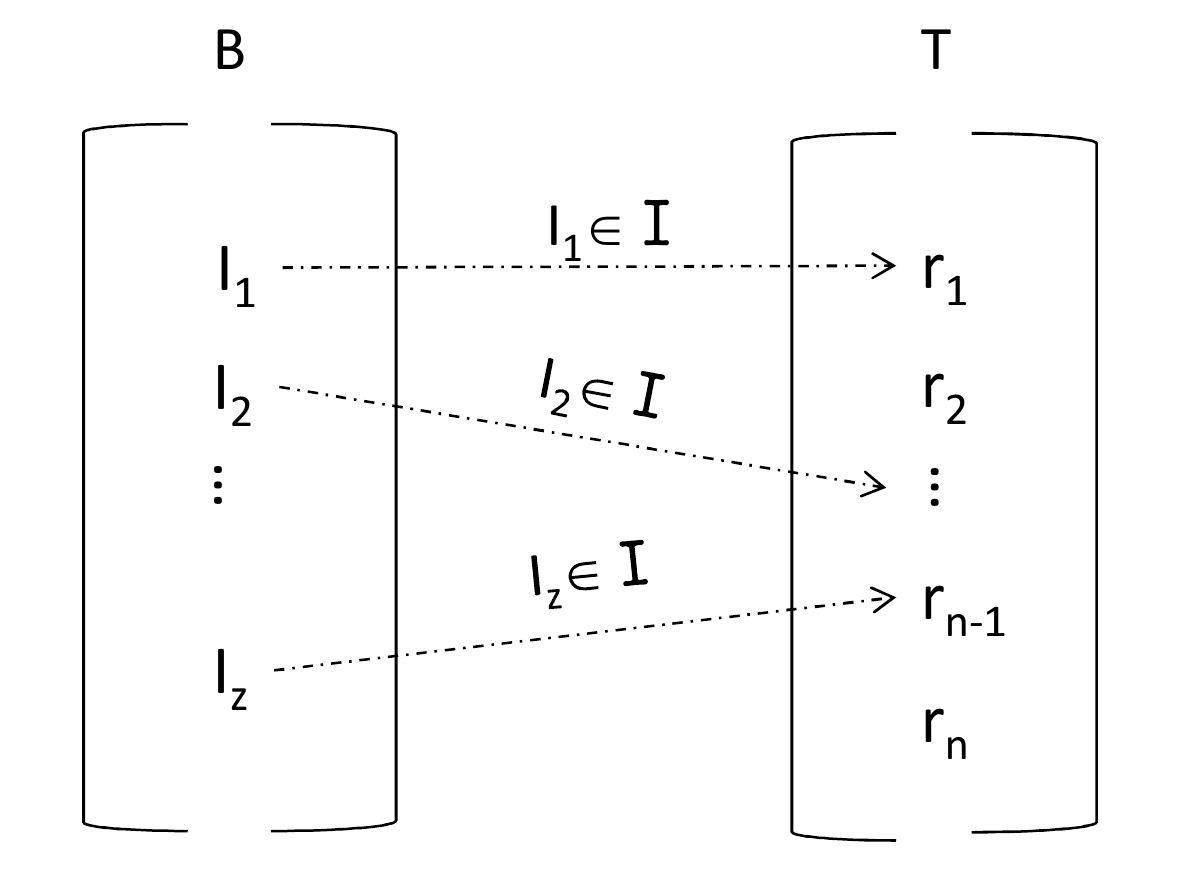}
	\label{fig:weak}
}
\caption{Associations between the background knowledge of attackers based on their category and the records in $\mathcal{T}$ }
\label{fig:attackers}\vspace*{-14pt}
\end{figure*}

\begin{description}
\item[Strong attackers] have a background knowledge $\mathcal{B}=\{I_1, ..., I_n\}$ such that $\forall I_i \in \mathcal{B}$,  $I_i$ is of size $m$,  $|I_i|=m$ and \textit{well picked}\footnote{More details on how these itemsets are picked will be provided in the experiments section} from the set of $m$-combinations $[r_i]_m$ of a record $r_i$, $I_i \in [r_i]_m$. For these attackers, as shown in Figure \ref{fig:strong}, there exists a bijective function that links itemsets in $\mathcal{B}$ to records in $\mathcal{T}$.
 
\item[Moderate attackers] have a background knowledge $\mathcal{B}=\{I_1, ..., I_z\}$, $z<n$  such that $\forall I_i \in \mathcal{B}$,  $I_i$ is of size $m$,  $|I_i|=m$ and \textit{chosen at random} from a strict subset $\mathcal{I}_v^-$ of the $m$-combinations $[r_v]_m$ of a record $r_v$, $I_i \in I_v^- | I_v^- \subset [r_v]_m$. For these attackers, as shown in Figure \ref{fig:moderate}, there exists a one-to-one injective function that links itemsets in $\mathcal{B}$ to records in $\mathcal{T}$. 

\item[Weak attackers] have a background knowledge  $\mathcal{B}=\{I_1, ..., I_z\}$, $z<n$ such that $\forall I_i \in \mathcal{B}$,  $I_i$ is of size $m$,  $|I_i|=m$ and \textit{chosen at random} from a set of association rules and frequent itemsets $\mathcal{I}$, $I_i \in I$,  that are not necessarily in $\mathcal{D}$. The goal of weak attackers is limited to reconstruct the disassociated dataset to its original form but they are unable to link records to individuals (see Figure \ref{fig:weak}).

\end{description}

\section{Privacy Evaluation in a Disassociated Dataset}\label{sec:deanon}



In this section, we define the cover problem and we demonstrate its repercussion on data privacy. More specifically, we evaluate the privacy breach that might occur in a disassociated dataset due to the cover problem. 

\subsection{Cover Problem}

A cover problem is defined by the ability to associate one-to-one or one-to-many items in two subsequent record chunks in the disassociated data. The target item from the first chunk has equal or higher support and we call it the covered item.  
Formally the cover problem is defined as follows:  

\begin{definition}[Cover Problem]
	Given a set of items in $R_{i_{C_{j-1}}}$  $(j \ge 2)$ that have a support greater than or equal to the support of an item $x_{i,j} \in R_{i_{C_j}}$, 
	$I_{i,j-1}=\{y: y \in  R_{i_{C_{j-1}}} $ and $s(y, R_{i_{C_{j-1}}}) \geq s(x_{i,j}, R_{i_{C_j}}) \} $,
	 we say that a cover problem exists if $\exists y_{i,j-1} \in I_{i,j-1}$ such that 
	$s(y_{i,j-1}, R_{i_{C_{j-1}}} )$\\$ = s(I_{i,j-1}, R_{i_{C_{j-1}}}) = \displaystyle \min_{\forall y \in {I}_{i,j-1}} s(y, R_{i_{C_{j-1}}})$. \end{definition}

\begin{figure*}[htb]
	\centering
	\subfloat[Cover problem example in $\mathcal{T}^*$]{
		\includegraphics[width=0.1\textwidth]{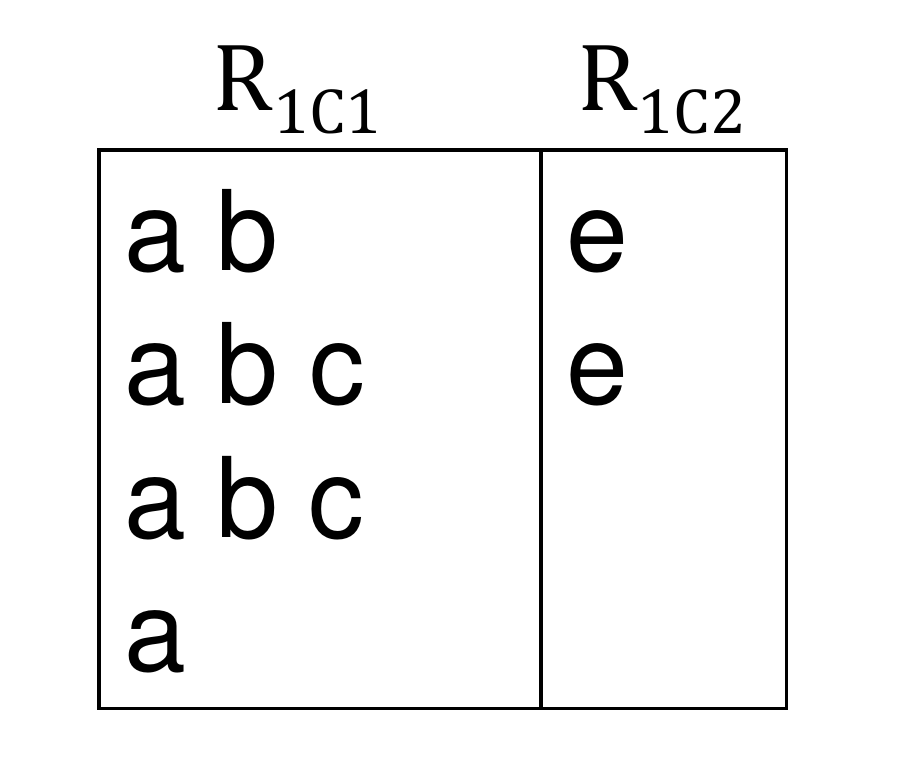}
\label{fig:coverProblem}
	}
\subfloat[Datasets reconstructed by the inverse transformation of $\mathcal{T}^*$]{
		\includegraphics[width=0.332\textwidth]{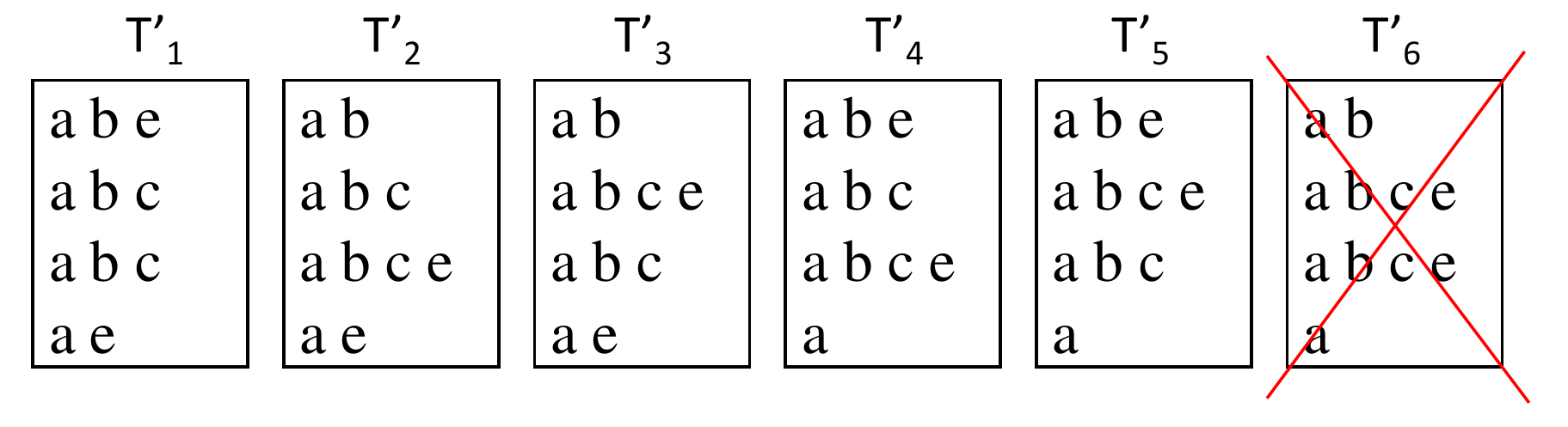}
	\label{fig:cattack-a}
}
\subfloat[Datasets in which items $e$ and $c$ are associated]{
		\includegraphics[width=0.24\textwidth]{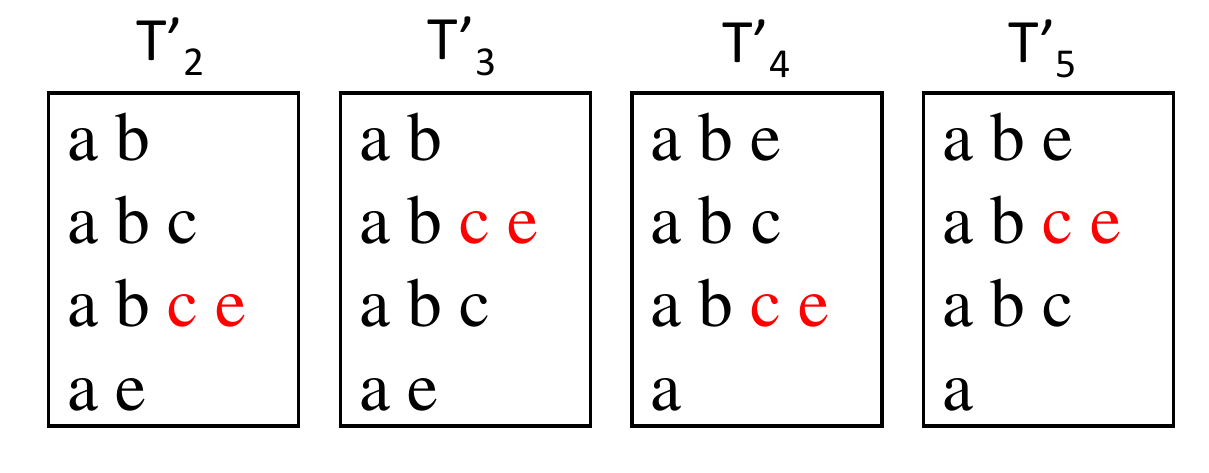}
	\label{fig:cattack-b}
}

\caption{Cover problem and privacy breach examples}
\label{fig:pbreach}\vspace*{-14pt}
\end{figure*}

To give an example of the cover problem, consider Figure \ref{fig:coverProblem}. The set of items in the previous record chunk $R_{1_{C_1}}$ having support greater than or equal to \textit{e} is  $I_{1,1}$ = \{$a, b, c$\}. 
The support of the itemset $I_{1,1}$ in $R_{1_{C_1}}$, which is $s(I_{1,1}, R_{1_{C_1}})$, is equal to 2. In turn, it is equal to the minimum support of the values in $I_{1,1}$, which, in our example, is $s(c, R_{1_{C_1}})$.
Therefore we say that the item \textit{c} is covered by the items \textit{a} and \textit{b}. 

Another example can be extracted from Figure \ref{fig:disasso-b}, the item \textit{Cancer} in $I_{1}$ is covered by the remainder of the items in $I_{1}$. In fact, all of the occurrences of the item \textit{Cancer} are associated with the items \textit{Treatment} and \textit{Oncologist} and therefore, by associating the item \textit{Side Effects} with \textit{Cancer}, \textit{Side Effects} will eventually be linked to both \textit{Treatment} and \textit{Oncologist}.
 
\subsection{Privacy Breach}\label{sec:pb}
Recall from Subsection \ref{sec:disass} and Definition 2, that the disassociation technique hides infrequent records; records that occur less than $k$ times in the original dataset for a given $m$ items, by 1) dividing them into $k$-anonymous sub-records in record chunks and 2) ensuring that all the records reconstructed by the inverse transformation are $k^m$-anonymous in at least one of the resulting datasets. 

Intuitively, a privacy breach occurs if an attacker is able to link $m$ items, which he/she already knows about an individual, to less than $k$ records in all the datasets reconstructed by the inverse transformation. More subtle is when these records contain the same set of items in all the reconstructed datasets thus, linking more than $m$ items to the individual or worse leading to a complete $de$-anonymization; linking, with certainty, the complete set of items to the individual. (Figure \ref{fig:cattack-b} is an example of such a $de$-anonymization. The attacker already knows that an individual has searched for items $e$ and $c$ and thus can relate him/her them to the same items $a$ and $b$ in all the reconstructed datasets).  

We will show in the following that this privacy breach might occur whenever the dataset is subject to a cover problem. Formally speaking:  
\begin{lemma}
	Let $T^*$ be a disassociated dataset subject to a cover problem and involving an item $x_{i,j} \in R_{i_{C_{j}}}$ and a covered item $y_{i,j-1} \in R_{i_{C_{j-1}}}$. We say that a privacy breach might occur if \{$x_{i,j}$, $y_{i,j-1}$\} $\subseteq I$ where $I\in \mathcal{B} $, the attacker's background knowledge.
	\end{lemma}
We show in what follows that an attacker is able to breach the privacy provided by the disassociation technique, if his/her background knowledge contains the items involved in the association that led to the cover problem.

\begin{proof} 

	  
	   The dataset $\mathcal{T}^*$ is subject to a cover problem and therefore, there exists an item  $x_{i,j} \in R_{i_{C_j}} (j \ge 2)$ that can be associated with a covered item $y_{i,j-1} \in  I_{i,j-1}$  where  $I_{i,j-1}=\{y: y \in R_{i_{C_{j-1}}} $ and $s(y, R_{i_{C_{j-1}}}) \geq s(x_{i,j}, R_{i_{C_j}})\}$ is the set of items in $ R_{i_{C_{j-1}}}$ having support greater than or equal to the support of $ x_{i,j}$.  
	 
	  Let us assume that the items $x_{i,j}$ and $y_{i,j-1}$ are associated together in $k$ records in at least one of the datasets reconstructed by the inverse transformation of $\mathcal{T}^*$.
	 Since $y_{i,j-1}$ is a covered item, $x_{i,j}$ will also be associated $k$ times with all the items covering $y_{i,j-1}$ in $I_{i,j-1}$ . While this is correct from a privacy perspective, it cannot be considered for disassociation. In fact, these items, $x_{i,j}$, $y_{i,j-1}$ and the covering items in $I_{i,j-1}$ are considered $k$-anonymous and therefore should have been allot to the record chunk $R_{i_{C_{j-1}}}$ according to disassociation\footnote{Vertical partitioning, creation of record chunks, preserves $k$-anonymous itemsets in the same record chunks.}. 
	  
	   
	  Now, if the attacker's background knowledge $\mathcal{B}$ consists of itemsets of size $m$ and there exists an itemset $I$ that contains both $x_{i,j}$ and $y_{i,j-1}$, \{$x_{i,j}$, $y_{i,j-1}$\} $\subseteq I$, a privacy breach will occur. In fact, these $m$ items will never be associated together in $k$ records in any of the datasets reconstructed by the inverse transformation of $\mathcal{T}^*$.\vspace{-7pt}
\end{proof}

Figure \ref{fig:cattack-a} describes six possible datasets, $\mathcal{T}^{'}_1$, ...,$\mathcal{T}^{'}_6$, reconstructed by the inverse transformation of the disassociated dataset shown in Figure \ref{fig:coverProblem}. In this example, only five datasets $\mathcal{T}^{'}_1$, ..., $\mathcal{T}^{'}_5$ are valid ones. $\mathcal{T}^{'}_6$ is omitted due to the cover problem and the knowledge of the disassociation algorithm. Given that the item $e$ is associated with the covered item $c$ is $k=2$ times, consequently $e$ is associated with $a$ and $b$ in $k$ records and thus, according to the disassociation technique, the item $e$ should have remained in the first record chunk $R_{1_{C_1}}$.  
 
If the attacker's background knowledge contains the items $c$ and $b$, he/she will be able to link every record containing $c$ and $b$ to less than $k$ records in the record chunk, thus breaching the privacy of disassociation (see Figure \ref{fig:cattack-b}).

This privacy breach is also detected in the example of Figure \ref{fig:disasso-b}. 
The item \textit{Cancer} in $R_{1_{C_1}}$, is one of the least frequent items in the set of items having support greater than or equal to the item \textit{Side Effects} in the record chunk $R_{1_{C_2}} , I_{1} = \{\textit{Oncologist, Cancer, Treatment}\}$. \textit{Cancer} is covered by the remainder of the items in $I_{1}$. If \textit{Side Effects} is associated with \textit{Cancer} in $k$ records, \textit{Side Effects} will be associated with \textit{Treatment} and \textit{Oncologist} in $k$ records or alternatively saying that the itemset \{ \textit{Side Effects}, \textit{Cancer}, \textit{Treatment} and \textit{Oncologist} \} is $k$-anonymous. As a consequence, according to the disassociation technique, these items should have belonged to the same record chunk which does not correspond to the disassociated dataset of Figure \ref{fig:disasso-b}.

\subsection{Quantitative Privacy Breach Detection Algorithm}

We present, in the following, the quantitative privacy breach detection algorithm to evaluate the privacy breach in a disassociated dataset based on the cover problems. 
The algorithm takes a disassociated dataset $\mathcal{T^*}$, and the attacker's background knowledge $\mathcal{B}$ that includes itemsets of size $m$ 
and 
evaluates 
the privacy breach represented by the number of vulnerable records in the disassociated dataset $\mathcal{T}^*$.
It iterates, in Step 3, over each cluster $R_{i}$ in $\mathcal{T}^*$ in an ascending order to evaluate, for each item $x_{i,j}$ in record chunk $R_{i_{C_j}}$, the number of privacy breaches in all the previous record chunks.
To do so, the algorithm 1) retrieves in Step 10 the itemset $I_{i,l}$ of items having support greater than or equal to the support of $x_{i,j}$ and 2) verifies in Step 11 whether $x_{i,j}$ and any item $y$ in $I_{i,l}$ is subject to a cover problem. 
A privacy breach is noted if both items $x_{i,j}$ and $y$ are included in any of the itemsets $I$ of the attackers' background knowledge $\mathcal{B}$.  The algorithm returns the total number of vulnerable records that is computed based on the sum of the maximum number of privacy breaches determined per item.

\begin{algorithm}[htp]
	\caption{Quantitative Privacy Breach Detection algorithm}
	\small

	\begin{algorithmic}[1]	\Require{ $\mathcal{T}^*, \mathcal{B}$}
	\Ensure {total}
		\State $max \gets 0$;
		\State $total \gets 0$;
		\For {each $R_{i} \in \mathcal{T}^*$}
				\State get $j$, the number of record chunks in $R_{i}$;
		\For {each $R_{i_{C_{j}}} \in R_{i}$ }

		\State $breach \gets 0$;
		\State $l \gets j-1$;
		\For {each item $x_{i,j} \in R_{i_{C_{j}}}$}
		\While {($l \geq 0$)}
		\State get $I_{i,l}$;

		\Statex	
		\Comment{$I_{i,l}=\{y : y \in R_{i_{C_l}}$ and $s(y, R_{i_{C_{l}}}) \geq s(x_{i,j}, R_{i_{C_j}}) \} $}

		\If{$ s(I_{i,l}, R_{i_{C_{l}}}) = \displaystyle \min_{\forall y \in {I}_{i,l}} s(y, R_{i_{C_{l}}})$ and $\{y_{i,l}, x_{i,j}\} \subseteq I $ }
		\Statex	
		\Comment {$I \in \mathcal{B}$}
		\State $breach++;$
		\EndIf
		\State $l--$;
		\EndWhile
		\EndFor
		\If {$breach > max$}
		\State $max \gets breach$;
		\EndIf
		\EndFor
		\State $total\gets total + max$;
		\EndFor 
		\State return $total$;
	\end{algorithmic}
\end{algorithm}
Although the proposed algorithm covers a broader aspect of the privacy breach by estimating the number of records that are vulnerable due to the cover problem, it can be extended to provide better insights on how many records are $de$-anonymized; linked to the individual with certainty. In fact, as discussed in Subsection \ref{sec:pb}, a privacy breach is found not only when completely $de$-anonymizing a record but also when an attacker is able to link more than $m$ items to the individual. This is noted in the algorithm every time the items involved in the association that led to the cover problem are part of the attacker's knowledge.

\section{Experiments} \label{sec:experim}

\begin{figure*}[htb]
	\centering
\subfloat[Strong attacker privacy breach evaluation]{
		\includegraphics[width=0.33\textwidth]{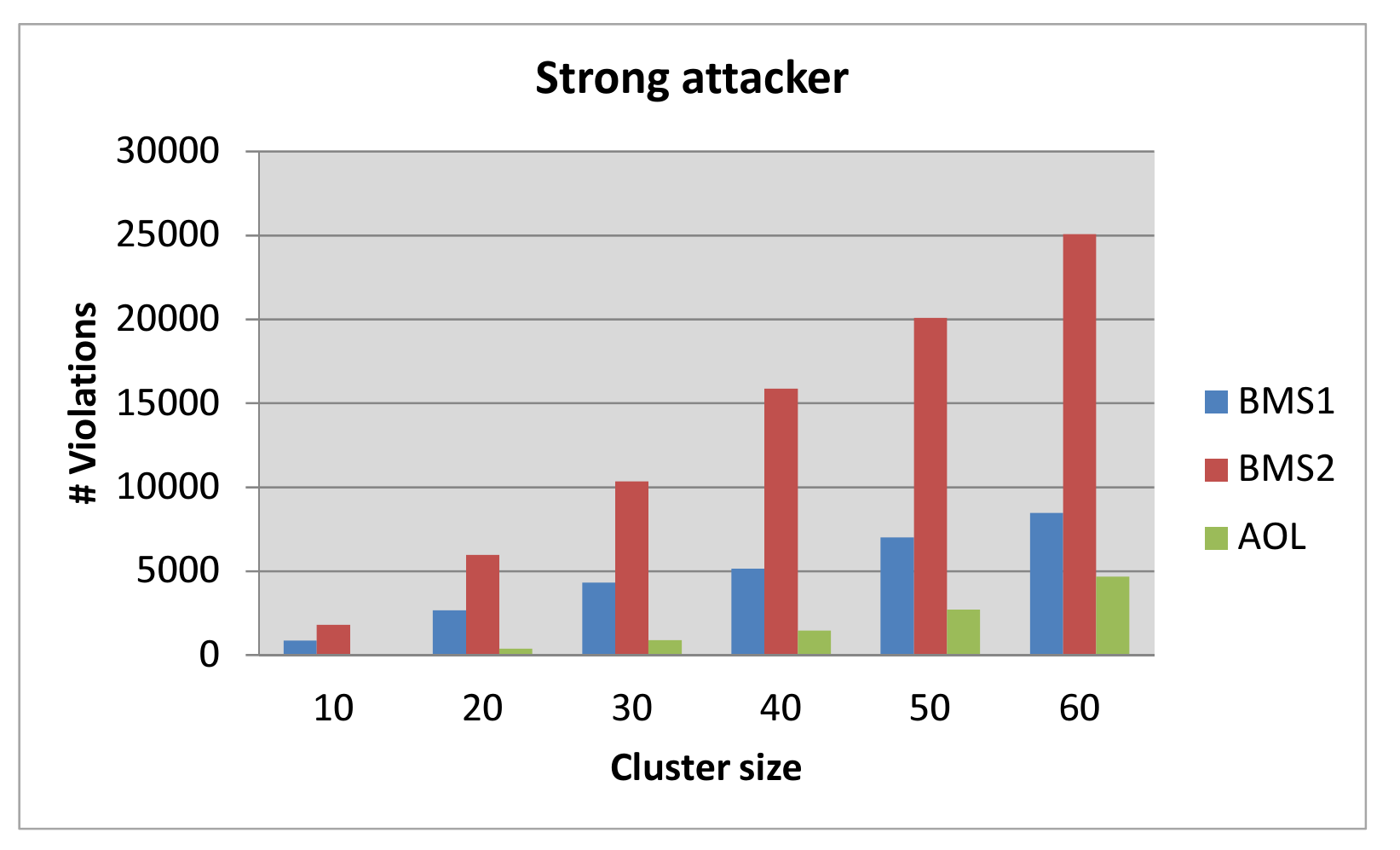}
	\label{fig:strong-a}
}
\subfloat[Moderate attacker privacy breach evaluation]{
		\includegraphics[width=0.33\textwidth]{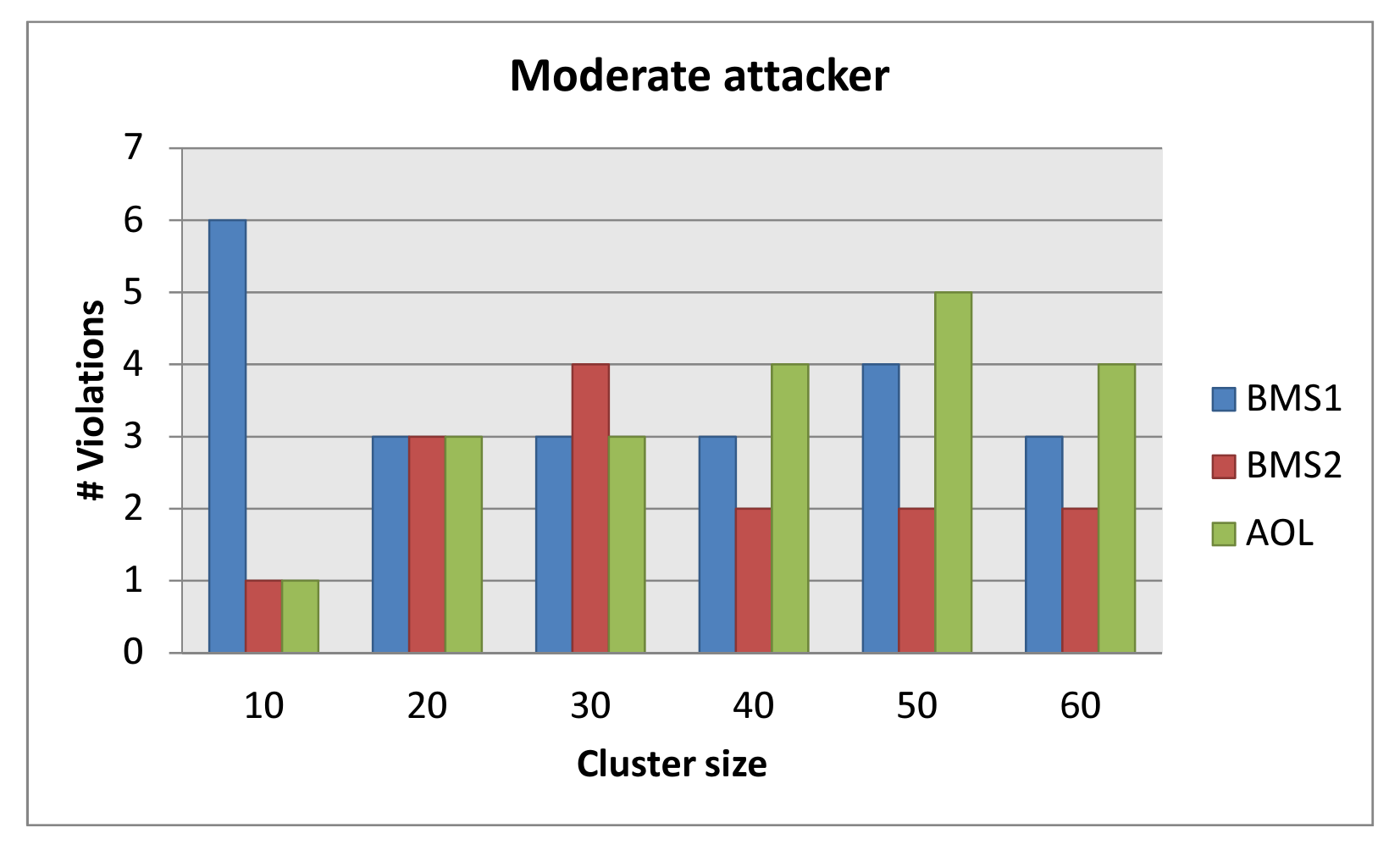}
	\label{fig:moderate-b}
}
\subfloat[Weak attacker privacy breach evaluation]{
		\includegraphics[width=0.33\textwidth]{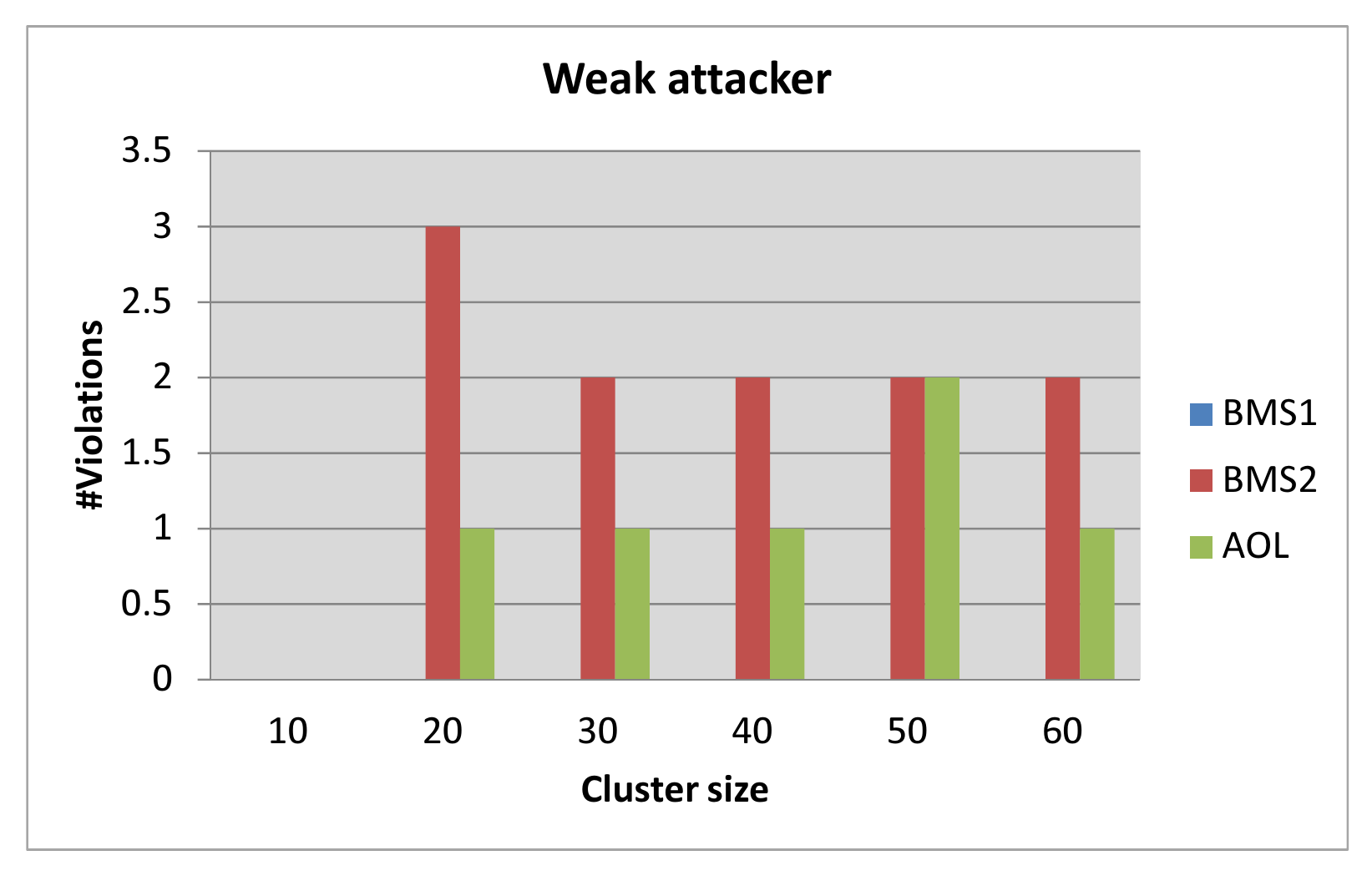}
	\label{fig:weak-b}
}\\

\caption{Privacy breach and performance evaluations }
\label{fig:pbreacheval}\vspace*{-10pt}
\end{figure*}

We conducted a series of experiments on real datasets described in Table \ref{tab:datasets}, the click-stream data, BMS-WebView-1 and BMS-WebView-2, and the AOL search query logs. 
We implemented all methods in JAVA and performed the experiments on an Intel Core i7-4702MQ CPU at 2.2GHz with 8GB of RAM. 
The goal of these experiments is to:
\begin{itemize}
\item Evaluate the privacy breach represented by the number of vulnerable records using the quantitative privacy breach detection algorithm on the aforementioned datasets with various attackers: strong, moderate, and weak,

\item  Evaluate the performance of the quantitative privacy breach detection algorithm.
\end{itemize} 

\begin{table}[htp]
\tiny
\centering
\begin{tabular}{ | l | c | r | }
  \hline
  Dataset & \# of distinct individuals & \# of distinct items \\ \hline
  AOL & 65517 & 1216655 \\ \hline
  BMS-WebView-1 & 59602 & 497 \\ \hline
  BMS-WebView-2 & 77512 & 3340 \\ \hline
\end{tabular}

\caption{Datasets properties}
\label{tab:datasets}
\end{table}

The tests were performed using three types of attackers, strong, moderate and weak, where each of which has a background knowledge simulated as follows:
\begin{description}
\item [Strong attacker]: the background knowledge of this attacker consists of itemsets of size 2 ($m=2$) picked from the $2$-combinations of each and every record in the dataset. Since it is a strong attacker, we assume that these itemsets contain the items that are involved in the association that will lead to a cover problem (i.e., well picked). Therefore, a privacy breach occurs whenever a cover problem is found.
\item [Moderate attacker]: the background knowledge of this attacker consists of itemsets of size 2 ($m=2$) chosen at random from the $2$-combinations of a strict subset of $\mathcal{D}$ (records in $\mathcal{T}$).
\item [Weak attacker]: to simulate the background knowledge of this attacker, we have chosen at random 10 items from the dataset (for each test) and another 10 items from wordnet \cite{miller95} and generated their $2$-combinations. This gave us a background knowledge $\mathcal{B}$ of size 190. 

\end{description}

\paragraph*{Privacy Breach Evaluation}
In this test, we evaluate the privacy breach caused by the cover problem using the quantitative privacy breach detection algorithm. Three test cases were studied. In each of the test cases, the algorithm evaluates the aforementioned datasets by varying the maximum cluster size with fixed $k=3$, $m=2$ according to the attacker's background knowledge. 
The results are shown in Figure \ref{fig:pbreacheval}. 
It is not surprising that for strong attackers the privacy breach is most explicit. In fact, strong attackers' background knowledge, consisting of all the itemsets of size $m=2$, are linked to each of the individuals in the dataset. Therefore, whenever a cover problem is found, a privacy breach is noted. 
In addition, we can notice that for the strong attacker as exhibited in the three datasets, the privacy breach increases linearly with the cluster size. Larger clusters include more cover problems and thus more privacy breaches. 
This is not reflected by moderate and weak attackers. These attackers have background knowledge generated from a strict subset of randomly chosen items from the dataset and other items from external sources. Moreover and unlike the strong attackers' background knowledge, moderate and weak attackers' background knowledge are linked to a subset of records in the datasets. This fact justifies the low number of violations found in the dataset for these attackers.
\begin{figure}

		\includegraphics[width=0.4	\textwidth]{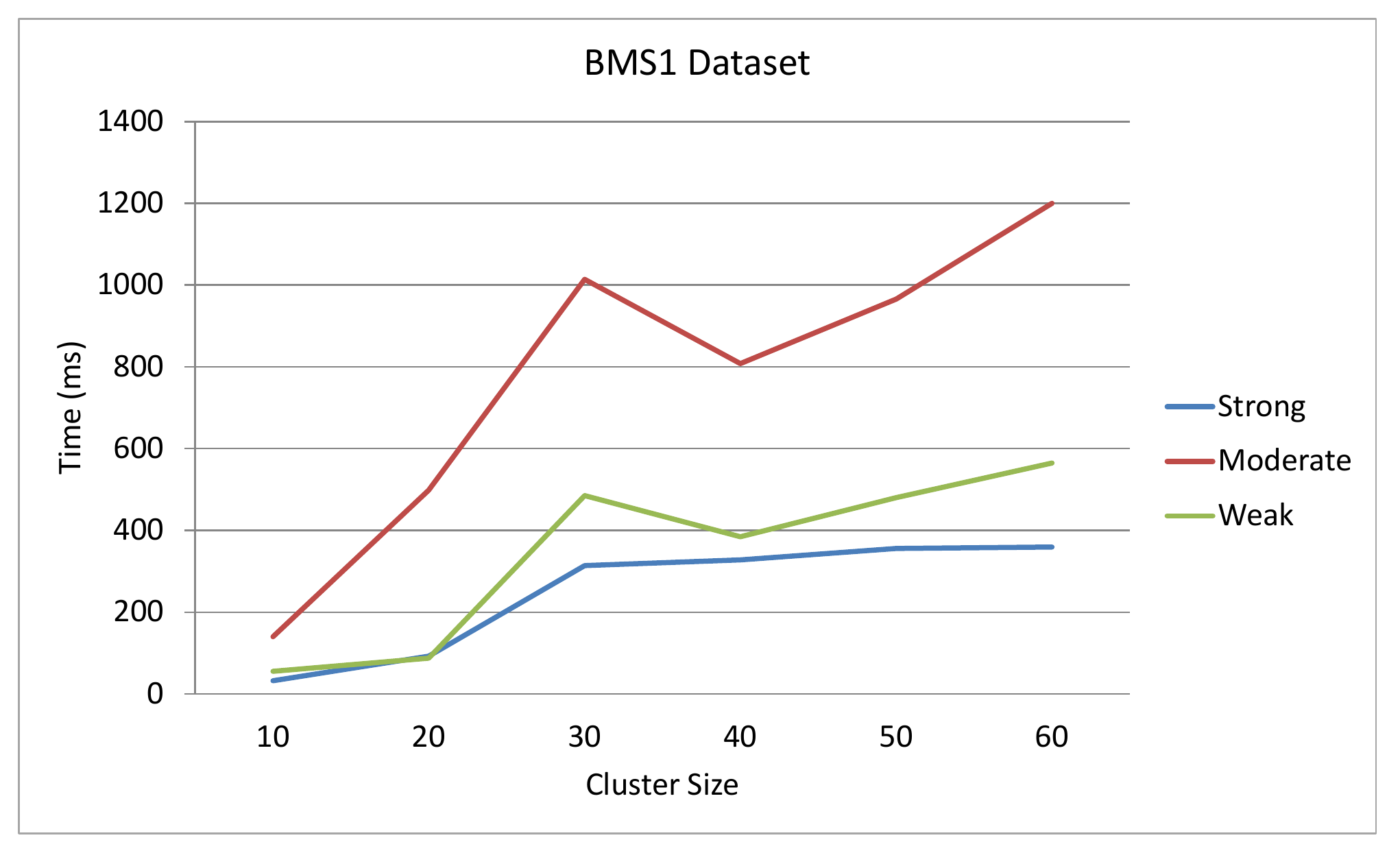}

\caption{Quantitative privacy breach detection algorithm performance evaluation}	\label{fig:perfEval}\vspace*{-10pt}
\end{figure}
\paragraph*{Performance Evaluation}

Here, we evaluate the performance of the quantitative privacy breach detection algorithm on the BMS-WebView-1 dataset for the three types of attackers; strong, moderate and weak. 
The results shown in Figure \ref{fig:perfEval} demonstrate that the algorithm performs in a generally stable fashion when increasing the cluster size. 
The best performance is obtained when dealing with strong attackers. The algorithm treats every cover problem as a privacy breach and therefore, unlike for moderate and weak attackers, there is no need to search for the items that are involved in the association that led to a cover problem in the background knowledge of the attackers.

\section{Conclusion} \label{sec:conclusion}

In this article, we have proposed a formal model of the set-valued data and the disassociation technique. The cover problem has been defined, while its effects on the disassociation of the dataset has been regarded. Such investigations have led to the quantitative privacy breach detection algorithm, whose efficiency has been studied. By this way, we have shown in what extent a disassociated dataset can be vulnerable. 
In the near future, we intend to develop partial suppression in disassociated dataset and evaluate the gain in utility that the disassociation provides.

\bibliographystyle{apalike}
\small
\bibliography{references}

\begin{thebibliography}{}

\bibitem[al~Bouna et~al., 2015a]{bechara14}
al~Bouna, B., Clifton, C., and Malluhi, Q.~M. (2015a).
\newblock Anonymizing transactional datasets.
\newblock {\em Journal of Computer Security}, 23(1):89--106.

\bibitem[al~Bouna et~al., 2015b]{bechara15}
al~Bouna, B., Clifton, C., and Malluhi, Q.~M. (2015b).
\newblock Efficient sanitization of unsafe data correlations.
\newblock In {\em Proceedings of the Workshops of the {EDBT/ICDT} 2015 Joint
  Conference (EDBT/ICDT), Brussels, Belgium, March 27th, 2015.}, pages
  278--285.

\bibitem[Barbaro and Zeller, 2006]{aol2006}
Barbaro, M. and Zeller, T. (2006).
\newblock A face is exposed for aol searcher no. 4417749.

\bibitem[Biskup et~al., 2011]{fragmentationInference}
Biskup, J., PreuB, M., and Wiese, L. (2011).
\newblock On the inference-proofness of database fragmentation satisfying
  confidentiality constraints.
\newblock In {\em Proceedings of the 14th Information Security Conference},
  Xian, China.

\bibitem[Ciriani et~al., 2010]{fragmentation}
Ciriani, V., Vimercati, S. D. C.~D., Foresti, S., Jajodia, S., Paraboschi, S.,
  and Samarati, P. (2010).
\newblock Combining fragmentation and encryption to protect privacy in data
  storage.
\newblock {\em ACM Trans. Inf. Syst. Secur.}, 13:22:1--22:33.

\bibitem[Cormode et~al., 2010]{cormodeMinimality}
Cormode, G., Li, N., Li, T., and Srivastava, D. (2010).
\newblock Minimizing minimality and maximizing utility: Analyzing method-based
  attacks on anonymized data.
\newblock In {\em Proceedings of the VLDB Endowment}, volume~3, pages
  1045--1056.

\bibitem[Dwork et~al., 2006]{Dwork2006}
Dwork, C., McSherry, F., Nissim, K., and Smith, A. (2006).
\newblock Calibrating noise to sensitivity in private data analysis.
\newblock In {\em Proceedings of the Third Conference on Theory of
  Cryptography}, TCC'06, pages 265--284, Berlin, Heidelberg. Springer-Verlag.

\bibitem[Fard and Wang, 2010]{fard2010effective}
Fard, A.~M. and Wang, K. (2010).
\newblock An effective clustering approach to web query log anonymization.
\newblock In {\em Security and Cryptography (SECRYPT), Proceedings of the 2010
  International Conference on}, pages 1--11. IEEE.

\bibitem[He and Naughton, 2009]{yeye2009}
He, Y. and Naughton, J.~F. (2009).
\newblock Anonymization of set-valued data via top-down, local generalization.
\newblock {\em Proc. VLDB Endow.}, 2(1):934--945.

\bibitem[Jia et~al., 2014]{Jia2014}
Jia, X., Pan, C., Xu, X., Zhu, K., and Lo, E. (2014).
\newblock ρ-uncertainty anonymization by partial suppression.
\newblock In Bhowmick, S., Dyreson, C., Jensen, C., Lee, M., Muliantara, A.,
  and Thalheim, B., editors, {\em Database Systems for Advanced Applications},
  volume 8422 of {\em Lecture Notes in Computer Science}, pages 188--202.
  Springer International Publishing.

\bibitem[Kifer, 2009]{definetti}
Kifer, D. (2009).
\newblock Attacks on privacy and definetti's theorem.
\newblock In {\em SIGMOD Conference}, pages 127--138.

\bibitem[Li et~al., 2012]{slicing}
Li, T., Li, N., Zhang, J., and Molloy, I. (2012).
\newblock Slicing: A new approach for privacy preserving data publishing.
\newblock {\em IEEE Trans. Knowl. Data Eng.}, 24(3):561--574.

\bibitem[Loukides et~al., 2014a]{loukidesLGT14}
Loukides, G., Liagouris, J., Gkoulalas{-}Divanis, A., and Terrovitis, M.
  (2014a).
\newblock Disassociation for electronic health record privacy.
\newblock {\em Journal of Biomedical Informatics}, 50:46--61.

\bibitem[Loukides et~al., 2014b]{Loukides2014}
Loukides, G., Liagouris, J., Gkoulalas-Divanis, A., and Terrovitis, M. (2014b).
\newblock Disassociation for electronic health record privacy.
\newblock {\em Journal of Biomedical Informatics}, 50(0):46 -- 61.
\newblock Special Issue on Informatics Methods in Medical Privacy.

\bibitem[Loukides et~al., 2015]{Loukides2015}
Loukides, G., Liagouris, J., Gkoulalas-Divanis, A., and Terrovitis, M. (2015).
\newblock Utility-constrained electronic health record data publishing through
  generalization and disassociation.
\newblock In Gkoulalas-Divanis, A. and Loukides, G., editors, {\em Medical Data
  Privacy Handbook}, pages 149--177. Springer International Publishing.

\bibitem[Machanavajjhala et~al., 2006]{ldiversity}
Machanavajjhala, A., Gehrke, J., Kifer, D., and Venkitasubramaniam, M. (2006).
\newblock $l$-diversity: Privacy beyond $k$-anonymity.
\newblock In {\em Proceedings of the 22nd IEEE International Conference on Data
  Engineering ({ICDE} 2006)}, Atlanta Georgia.

\bibitem[Miller, 1995]{miller95}
Miller, G.~A. (1995).
\newblock Wordnet: A lexical database for english.
\newblock {\em Commun. ACM}, 38(11):39--41.

\bibitem[Ressel, 1985]{definettitheorem}
Ressel, P. (1985).
\newblock De {F}inetti-type theorems: an analytical approach.
\newblock {\em Ann. Probab.}, 13(3):898--922.

\bibitem[Samarati, 2001]{kanon_defn}
Samarati, P. (2001).
\newblock Protecting respondents' identities in microdata release.
\newblock {\em IEEE Trans. Knowl. Data Eng.}, 13(6):1010--1027.

\bibitem[Sweeney, 2001]{Sweeney01computationaldisclosure}
Sweeney, L. (2001).
\newblock Computational disclosure control - a primer on data privacy
  protection.
\newblock Technical report, Massachusetts Institute of Technology.

\bibitem[Sweeney, 2002]{kanon_sweeney}
Sweeney, L. (2002).
\newblock k-anonymity: a model for protecting privacy.
\newblock {\em International Journal on Uncertainty, Fuzziness and
  Knowledge-based Systems}, 10(5):557--570.

\bibitem[Terrovitis et~al., 2008]{terrovitis08}
Terrovitis, M., Mamoulis, N., and Kalnis, P. (2008).
\newblock Privacy-preserving anonymization of set-valued data.
\newblock {\em {PVLDB}}, 1(1):115--125.

\bibitem[Terrovitis et~al., 2012]{terrovitis2012}
Terrovitis, M., Mamoulis, N., Liagouris, J., and Skiadopoulos, S. (2012).
\newblock Privacy preservation by disassociation.
\newblock {\em Proc. VLDB Endow.}, 5(10):944--955.

\bibitem[Wong et~al., 2007]{wong2007}
Wong, R. C.-W., Fu, A. W.-C., Wang, K., and Pei, J. (2007).
\newblock Minimality attack in privacy preserving data publishing.
\newblock In {\em VLDB}, pages 543--554.

\bibitem[Wong et~al., 2011]{correlation}
Wong, R. C.-W., Fu, A. W.-C., Wang, K., Yu, P.~S., and Pei, J. (2011).
\newblock Can the utility of anonymized data be used for privacy breaches?
\newblock {\em ACM Trans. Knowl. Discov. Data}, 5(3):16:1--16:24.

\bibitem[Xiao and Tao, 2006]{anatomy}
Xiao, X. and Tao, Y. (2006).
\newblock Anatomy: Simple and effective privacy preservation.
\newblock In {\em Proceedings of 32nd International Conference on Very Large
  Data Bases (VLDB 2006)}, Seoul, Korea.

\end{thebibliography}

\end{document}